\documentclass{aastex}
\usepackage{spr-astr-addons}
\usepackage{url}
\usepackage[dvips]{color}
\usepackage{epsfig}
\usepackage{amsmath}
\usepackage{graphicx}
\RequirePackage{color}

\begin{document}
\title{FRW Bulk Viscous Cosmology with Modified Chaplygin Gas in Flat Space}
\author{H. Saadat} \and \author{B. Pourhassan}
\affil{Department of Physics, Sepidan Branch, Islamic Azad University, Sepidan, Iran, P. O. Box 71555-477.\\
hsaadat2002@yahoo.com}

\begin{abstract}
In this paper we study FRW bulk viscous cosmology in presence of modified Chaplygin gas. We write modified Friedmann equations due to bulk viscosity and
Chaplygin gas and obtain time-dependent energy density for the special case of flat space.
\end{abstract}
\section{Introduction}
It is found that our universe expands with acceleration (Riess et al. 1998; Perlmutter et al. 1999; Knop et al. 2003; Riess et al. 2004; Bennet et al.
2003). The accelerating expansion of the universe may be explained in context of the dark energy (Bamba et al. 2012). Due to negative pressure, the
simplest way for modeling the dark energy is the Einstein's cosmological constant. On the other hand the study of the cosmological constant is one of the
important subject in the theoretical and experimental physics (Weinberg 1989; Padmanabhan 2003; Peebles and Ratra 2003; Nobbenhuis 2006). Another candidate
for the dark energy is scalar-field dark energy model (Peebles and Ratra 1988; Ratra and Peebles 1988; Turner and White 1997; Caldwell 2002; Sen 2002; Feng
et al. 2005;  Guo et al. 2005; Wei 2005; Wei et al. 2007). However presence of a scalar field is not only requirement of the transition from a universe
filled with matter to an exponentially expanding universe. Therefore, Chaplygin gas (Setare 2007) used as an exotic type of fluid, which is based on the
recent observational fact that the equation of state parameter for dark energy can be less than $-1$. Important picture of Chaplygin gas may seen in
context of holography (Setare 2007; Setare 2009) where a correspondence between the holographic dark energy and Chaplygin gas energy density proposed.\\
On the other hand we know that the viscosity plays an important role in the cosmology (Singh and Devi 2011; Singh and Kale 2011; Setare and Sheykhi 2010;
Misner 1969). In another word, the presence of viscosity in the fluid introduces many interesting pictures in the dynamics of homogeneous cosmological
models, which is used to study the evolution of universe. Already (Chatterjee and Bhui 1990) the exact solutions of the field equations for a
five-dimensional space-time with viscous fluid obtained. In another work (Chatterjee and Bhui 1990) a cosmological model with viscous fluid in
higher-dimensional space-time constructed. In the interesting work (Singh et al. 2004) the exact solutions of the field equations for a five-dimensional
cosmological model with variable bulk viscosity obtained. The isotropic homogeneous spatially flat cosmological model with bulk viscous fluid is also
constructed (Murphy 1973). Then the bulk viscous cosmological models with constant bulk viscosity coefficient constructed  (Bali and Dave 2002). In the
recent work (Katore et al. 2011) the FRW bulk viscous cosmology considered and bulk viscous coefficient obtained in the flat space, and then extended to
non-flat space (Saadat 2012). In this work we consider both bulk viscous effect and Chaplygin gas in FRW cosmology in flat space. Indeed we modify
Friedmann equation due to Chaplygin gas which has bulk viscosity. We should here note that, this modified theory may be stable and energy-momentum
conserved. Stability of this theory is discussed and appropriate condition to have stable theory is obtained.
\section{Friedmann equations}
The Friedmann-Robertson-Walker (FRW) universe in four-dimensional space-time is described by the following metric (Saha et al. 2012; Jamil et al. 2012),
\begin{equation}\label{s1}
ds^2=-dt^2+a^{2}(t)\left(\frac{dr^2}{1-kr^{2}}+r^{2}d\Omega^{2}\right),
\end{equation}
where $d\Omega^{2}=d\theta^{2}+\sin^{2}\theta d\phi^{2}$, and $a(t)$ represents the scale factor. The $\theta$ and $\phi$ parameters are the usual
azimuthal and polar angles of spherical coordinates, with $0\leq\theta\leq\pi$ and $0\leq\phi<2\pi$. The coordinates ($t, r, \theta, \phi$) are called
co-moving coordinates. Also, constant $k$ denotes the curvature of the space. In this paper we consider the case of $k=0$ only, which is corresponding to
flat space. In that case the Einstein equation is given by,
\begin{equation}\label{s2}
R_{\mu\nu}-\frac{1}{2}g_{\mu\nu}R=T_{\mu\nu}+g_{\mu\nu}\Lambda,
\end{equation}
where we assumed $c=1$ and $8\pi G = 1$. Also the energy-momentum tensor corresponding to the bulk viscous fluid and modified Chaplygin gas (Benaoum 2002;
Debnath et al. 2004; Xu et al. 2012; Bandyopadhyay 2012; Rudra et al. 2012;  Rudra 2012) is given by the following relation,
\begin{equation}\label{s3}
T_{\mu\nu}=(\rho+\bar{p})u_{\mu}u_{\nu}-\bar{p}g_{\mu\nu},
\end{equation}
where $\rho$ is the energy density and $u^{\mu}$ is the velocity vector with normalization condition $u^{\mu}u_{\nu}=-1$. Also, the total pressure and the
proper pressure involve bulk viscosity coefficient $\zeta$ and Hubble expansion parameter $H=\dot{a}/a$ are given by the following equations,
\begin{equation}\label{s4}
\bar{p}=p-3\zeta H,
\end{equation}
and
\begin{equation}\label{s5}
p=\gamma\rho-\frac{B}{\rho^{\alpha}},
\end{equation}
with $B>0$ and $0<\alpha\leq1$. The equation of state $\gamma$ is one of the most important quantity to describe the features of dark energy models. It is
clear that the parameter $\zeta$ shows bulk viscosity and $B$ shows effect of Chaplygin gas. Already (Mazumder et al. 2012) the dynamics of FRW cosmology
with modified Chaplygin gas as the matter formulated. Then the nature of the critical points are studied by evaluating the eigenvalues of the linearized
Jacobi matrix for the special case of $\alpha=0.6$. In this paper
we consider special case with $\alpha=0.5$ and extend the previous work (Mazumder et al. 2012) to including bulk viscous coefficient.\\
In that case the Friedmann equations are given by,
\begin{equation}\label{s6}
(\frac{\dot{a}}{a})^{2}=\frac{\rho}{3},
\end{equation}
and
\begin{equation}\label{s7}
2\frac{\ddot{a}}{a}+(\frac{\dot{a}}{a})^{2}=-\bar{p},
\end{equation}
where dot denotes derivative with respect to cosmic time $t$. The energy-momentum conservation law obtained as the following,
\begin{equation}\label{s8}
\dot{\rho}+3\frac{\dot{a}}{a}(\rho+\bar{p})=0.
\end{equation}
In the next section we try to obtain time-dependent density by using above equations.
\section{Time-dependent density}
Using the equations (4), (5) and (6) in the conservation relation (8) we have,
\begin{equation}\label{s9}
\dot{\rho}+\sqrt{3}(\gamma+1)\rho^{\frac{3}{2}}-3\zeta\rho-\sqrt{3}B=0.
\end{equation}
If we set $\zeta=0$, then one can extract energy density depend on scale factor (Mazumder et al. 2012),
\begin{equation}\label{s10}
\rho(a)=\left[\frac{1}{\gamma+1}(B+\frac{c}{\sqrt{a^{9(\gamma+1)}}})\right]^{\frac{2}{3}},
\end{equation}
where $c$ is an integration constant. Here we also consider bulk viscous coefficient and would like to obtain energy density depend on time. In order to
solve equation (9) we use the following ansatz,
\begin{equation}\label{s11}
\rho=\frac{A}{t^{2}}+\frac{E}{t}+ht+Ce^{bt},
\end{equation}
where constants $A$, $E$, $h$, $C$ and $b$ should be determined. Substituting relation (11) in the equation (9) gives us the following coefficients,
\begin{equation}\label{s12}
h=\sqrt{3}B,
\end{equation}
\begin{equation}\label{s13}
A=\frac{4}{3(\gamma+1)^{2}},
\end{equation}
\begin{equation}\label{s14}
E=\frac{2\zeta}{(\gamma+1)^{2}},
\end{equation}
\begin{equation}\label{s15}
C=\frac{(\gamma+1)^{2}}{4}\left[\frac{8\sqrt{3}\zeta^{2}}{(\gamma+1)^{3}}-\frac{3(\gamma+1)^{4}}{16\zeta^{2}}\right],
\end{equation}
\begin{equation}\label{s16}
b=\frac{\zeta\left[\sqrt{3}\zeta(\gamma+1)(B(\gamma+1)-\frac{9}{2}\zeta^{3})
+\frac{27}{32}(\gamma+\frac{1}{8})+\frac{9}{16}\zeta^{4}+\mathcal{O}(\gamma^{n})\right]}{8(\gamma+1)(\sqrt{3}\zeta^{4}-\frac{3}{128}(\gamma+1)^{7})},
\end{equation}
where,
\begin{equation}\label{s17}
\mathcal{O}(\gamma^{n})\equiv\frac{189}{64}\gamma^{2}+\frac{189}{32}\gamma^{3}+\frac{945}{128}\gamma^{4}+\frac{189}{32}\gamma^{5}+\frac{189}{64}\gamma^{6}
+\frac{27}{32}\gamma^{7}+\frac{27}{256}\gamma^{8}.
\end{equation}
If we neglect both bulk viscosity and presence of Chaplygin gas then,
\begin{equation}\label{s18}
\rho=\frac{4}{3(\gamma+1)^{2}t^{2}},
\end{equation}
which is agree with results of previous works [Saadat 2012; Mazumder et al. 2012) where $\rho\propto t^{-2}$ established. On the other hand for the large
bulk viscosity coefficient one can find that $b<0$ and hence $\rho\propto\zeta/t$ obtained. Also for the case of infinitesimal $\zeta$ one can obtain
constant negative energy density. It is interesting to check late time behavior of density. In that case the last term of the equation (11) is dominant so
one can say $\rho\sim C exp(bt)$. In the general case, equation (11) with coefficients (12)-(16) tells us that the energy density is decreasing function of
time. Such behavior happen for the Hubble expansion parameter which is discussed in the next section.
\section{Hubble and deceleration parameters}
By using time-dependent density in the relation (6) one can obtain Hubble expansion parameter. In that case we draw plot of Hubble expansion parameter in
the Fig. 1 for $\gamma\simeq1/3$. In that case the modified Chaplygin gas model describes the evolution of the universe from the radiation regime to the
$\Lambda$-cold dark matter scenario, where the fluid behaves as a cosmological constant, so there is an accelerated expansion of the universe.

\begin{figure}[th]
\begin{center}
\includegraphics[scale=.3]{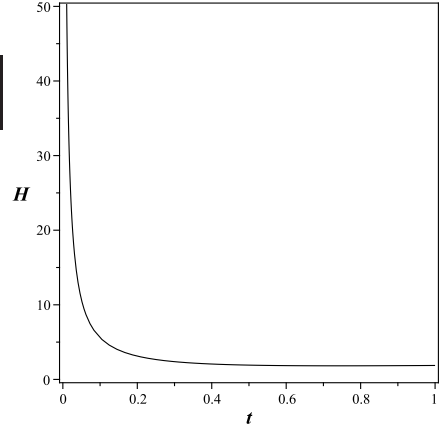} \caption{Hubble expansion parameter in terms of time for $B=3.4$, $\zeta=1$, and $\gamma=0.3$}
\end{center}
\end{figure}

It is possible to study deceleration parameter of this theory which obtained by the following relation,
\begin{equation}\label{s19}
q=-(1+\frac{\dot{H}}{H^{2}}).
\end{equation}

Numerically, we draw deceleration parameter in terms of time in the Fig. 2. It shows that the deceleration parameter yields to -1 at the late time. In the
case of $\zeta=0.2$ there is a maximum value of the deceleration parameter at $t\sim0.35$. In other cases time-dependent of the deceleration parameter is
completely decreasing.

\begin{figure}[th]
\begin{center}
\includegraphics[scale=.3]{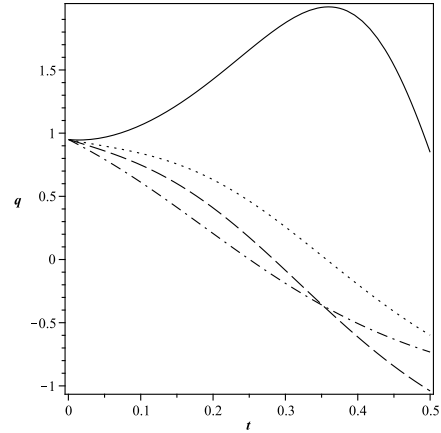} \caption{Deceleration parameter in terms of time for $B=3.4$ and $\gamma=0.3$. Solid, dotted, dashed and dash dotted lines represent
$\zeta=0.2, 0.4, 0.6, 1$ respectively.}
\end{center}
\end{figure}

\section{Stability}
It is important to investigate stability of this theory. There are several ways to do this. We use speed of sound in viscous fluid to study stability of
our system (Setare 2007; Sadeghi et al. 2010). In that case there is the following condition to have stable theory,
\begin{equation}\label{s20}
C_{s}^{2}=\frac{d\bar{p}}{d\rho}\geq0.
\end{equation}
We should use equations (4), (5) and (11) to satisfy the relation (20).  In the Fig. 3 we draw plot of $C_{s}^{2}$ for selected value of $\zeta$. It shows
that stability of theory is depend on viscosity. We find that for $\zeta<0.65$ theory is completely stable. On the other hand, in the case of $\zeta>0.65$
theory has unstable region. However at the late time, theory is completely stable and sound speed yields to constant value.

\begin{figure}[th]
\begin{center}
\includegraphics[scale=.4]{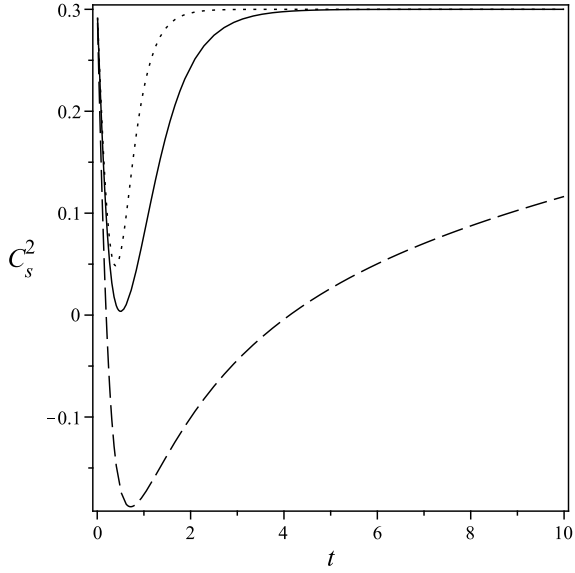} \caption{Square of sound speed in terms of time for $B=3.4$ and $\gamma=0.3$. Solid, dotted, and dashed lines represent
$\zeta=0.65, 0.6, 1$ respectively.}
\end{center}
\end{figure}

\section{Conclusion}
In this work we studied the FRW bulk viscous cosmology with modified Chaplygin gas as the matter contained. We obtained the modified Friedmann equations
due to bulk viscous and Chaplygin gas coefficients. Then, we tried to solve equations and found time-dependent energy density. Therefore, we could extract
Hubble expansion and deceleration parameters. We studied stability of theory and found that stability of system strongly is depend on viscosity
coefficient. However at the late time theory is stable and speed of sound has constant real value. For the future work it is possible to repeat calculation
of this paper for the case of arbitrary $\alpha$ or non-flat universe where $k\neq0$. It is also interesting to study thermodynamics of Chaplygin gas with
bulk viscosity similar to the recent work (Setare and Sheykhi 2010).


\begin{thebibliography}{11}
\bibitem{P1}
A.G. Riess et al., Astron. J. 116, 1009 (1998)
\bibitem{P2}
S. Perlmutter et al., Astrophys. J. 517, 565 (1999)
\bibitem{P3}
R.A. Knop et al., Ap. J. 598, 102 (2003)
\bibitem{P4}
A.G. Riess et al., Ap. J. 607, 665 (2004)
\bibitem{P5}
C.L. Bennet et al., Ap. J. Suppl. Ser. 148, 1 (2003)
\bibitem{P6}
K. Bamba, S. Capozziello, S. Nojiri, S. D. Odintsov, "Dark energy cosmology: the equivalent description via different theoretical models and cosmography
tests", Astrophys Space Sci DOI 10.1007/s10509-012-1181-8.
\bibitem{P7}
S. Weinberg, Rev. Mod. Phys. 61, 1 (1989). Sean M. Carroll, Living Rev. Relativity, 3, 1 (2001)
\bibitem{P8}
T. Padmanabhan, Phys. Rep. 380, 235 (2003)
\bibitem{P9}
P. J. E. Peebles, B. Ratra, Rev. Mod. Phys. 75, 559 (2003)
\bibitem{P10}
S. Nobbenhuis, Found. Phys. 36, 613 (2006), [arXiv:gr-qc/0411093]
\bibitem{P11}
P. J. E. Peebles and B. Ratra, Astrophys. J. 325, L17 (1988)
\bibitem{P12}
B. Ratra and P. J. E. Peebles, Phys. Rev. D 37, 3406 (1988)
\bibitem{P13}
M. S. Turner and M. J. White, Phys. Rev. D 56, 4439 (1997), [astro-ph/9701138]
\bibitem{P14}
R. R. Caldwell, Phys. Lett. B 545, 23 (2002), [astro-ph/9908168]
\bibitem{P15}
A. Sen, JHEP 0207, 065 (2002) [hep-th/0203265]
\bibitem{P16}
B. Feng, X. L. Wang and X. M. Zhang, Phys. Lett. B 607, 35 (2005), [astro-ph/0404224]
\bibitem{P17}
Z. K. Guo, Y. S. Piao, X. M. Zhang and Y. Z. Zhang, Phys. Lett. B 608, 177 (2005), [astro-ph/0410654]
\bibitem{P18}
H. Wei, and R.G. Cai, Phys. Rev. D72, 123507 (2005)
\bibitem{P19}
H. Wei, N.N. Tang, and R.G. Cai, Phys. Rev. D75, 043009 (2007)
\bibitem{P20}
M. R. Setare, Eur. Phys. J. C52, 689-692 (2007)
\bibitem{P21}
M. R. Setare, Phys. Lett. B648, 329-332 (2007)
\bibitem{P22}
M. R. Setare, Int. J. Mod. Phys. D18, 419-427 (2009)
\bibitem{P23}
N. Ibotombi Singh, S. Romaleima Devi, "A new class of bulk viscous FRW cosmological models in a scale covariant theory of gravitation", Astrophys Space Sci
(2011) 334:231–236.
\bibitem{P24}
G.P. Singh, A.Y. Kale, "Anisotropic bulk viscous cosmological models with particle creation", Astrophys Space Sci (2011) 331: 207–219.
\bibitem{P25}
M. R. Setare and A. Sheykhi, Int. J. Mod. Phys. D19, No. 2, 171 (2010)
\bibitem{P26}
C. V. Misner, Astrophys. J. 151, 431 (1969)
\bibitem{P27}
Chatterjee, S., Bhui, B.: Astrophys. Space Sci. 167, 61 (1990)
\bibitem{P28}
Banergee, A., Bhui, B., Chatterjee, S.: Astrophys. J. 358, 187 (1990)
\bibitem{P29}
Singh, G.P., Deshpande, R.V., Singh, T.: Pramana—J. Phys. 63, 937 (2004)
\bibitem{P30}
Murphy, G.L.: Phys. Rev. D 8, 4231 (1973)
\bibitem{P31}
Bali, R., Dave, S.: Astrophys. Space Sci. 282, 461 (2002)
\bibitem{P32}
S.D. Katore, A.Y. Shaikh, D.V. Kapse, S.A. Bhaskar, "FRW Bulk Viscous Cosmology in Multi Dimensional Space-Time", Int J Theor Phys (2011) 50:2644–2654
\bibitem{P33}
Hassan Saadat, "FRW bulk viscous cosmology in non-flat universe", Int J Theor Phys (2012) 51:1317–1322
\bibitem{P34}
B, Saha, H. Amirhashchi, A. Pradhan, "Two-fluid scenario for dark energy models in an FRW universe-revisited", Astrophys Space Sci DOI
10.1007/s10509-012-1155-x
\bibitem{P35}
M. Jamil, D. Momeni, N.S. Serikbayev, R. Myrzakulov, "FRW and Bianchi type I cosmology of f-essence", Astrophys Space Sci (2012) 339:37–43
\bibitem{P36}
Benaoum, H.B.: arXiv:hep-th/0205140 (2002)
\bibitem{P37}
Debnath, U., Banerjee, A., Chakraborty, S.: Class. Quantum Gravity 21, 5609 (2004)
\bibitem{P38}
Y.D. Xu, Z.G. Huang, X.H. Zhai, "A new type of interaction between generalized Chaplygin gas and dark matter", Astrophys Space Sci (2012) 339:31–36
\bibitem{P39}
Tanwi Bandyopadhyay, "Thermodynamics of Gauss-Bonnet brane with modified Chaplygin gas", Astrophys Space Sci DOI 10.1007/s10509-012-1115-5
\bibitem{P40}
P. Rudra, U. Debnath, R. Biswas, "Dynamics of modified Chaplygin gas in brane world scenario: phase plane analysis", Astrophys Space Sci (2012) 339:53–64
\bibitem{P41}
Prabir Rudra, "Dynamics of interacting generalized cosmic Chaplygin gas in brane-world scenario", Astrophys Space Sci DOI 10.1007/s10509-012-1198-z
\bibitem{P42}
N. Mazumder, R. Biswas, S. Chakraborty, "FRW Cosmological Model with Modified Chaplygin Gas and Dynamical System", Int J Theor Phys DOI
10.1007/s10773-012-1150-6
\bibitem{P43}
M. R. Setare, Phys. Lett. B654, 1-6 (2007)
\bibitem{P44}
J. Sadeghi, M. R. Setare, A. R. Amani and S. M. Noorbakhsh, "Bouncing Universe and Reconstructing Vector Field", [arXiv:1001.4682 [hep-th]] (2010)
\end{thebibliography}
\end{document}